\documentclass[10pt,twocolumn,prb,aps,amsmath,amssymb,superscriptaddress,longbibliography]{revtex4-2}

\usepackage{graphicx}
\usepackage{dcolumn}
\usepackage{bm}
\usepackage{float}
\usepackage[colorlinks=true,linkcolor=blue,citecolor=blue,urlcolor=blue]{hyperref}
\usepackage{natbib}

\begin{document}

\title{The Hidden Cost of Two-Dimensional TEM Reconstruction of
Magnetic Nanoparticle Morphology}



\author{Iago L\'opez-V\'azquez}
\thanks{Corresponding author: iago.lopez@usc.es}
\affiliation{Dpt. de Física Aplicada, Universidade de Santiago de Compostela, 15782 Santiago de Compostela, Galicia, Spain}
\affiliation{Instituto de Materiais (iMATUS), Universidade de Santiago de Compostela, 15782 Santiago de Compostela, Galicia, Spain}

\author{Amira P\'aez}
\email{ami@unizar.es}
\affiliation{Instituto de Nanociencia y Materiales de Arag\'on (INMA, UNIZAR-CSIC), Zaragoza, Spain}

\author{Yilian Fern\'andez-Afonso}
\email{y.f.afonso@csic.es}
\affiliation{Instituto de Ciencia de Materiales de Madrid (ICMM, CSIC), C/Sor Juana In\'es de la Cruz 3,28049, Madrid, Spain}%

\author{Roy W. Chantrell}
\email{roy.chantrell@york.ac.uk}
\affiliation{School of Physics, Engineering and Technology, University of York, York, United Kingdom}%

\author{Thomas S. van Zanten}
\email{t.vanzanten@csic.es}
\affiliation{Instituto de Nanociencia y Materiales de Arag\'on
(INMA, UNIZAR-CSIC), Zaragoza, Spain}
\affiliation{Fundación Agencia Aragonesa para la Investigación y el Desarrollo (ARAID), Zaragoza, Spain}

\author{Sergiu Ruta}
\email{sergiu.ruta@shu.ac.uk}
\affiliation{College of Business, Technology and Engineering, Sheffield Hallam University, Sheffield, United kingdom}

\author{Luc\'ia Guti\'errez}
\affiliation{Instituto de Nanociencia y Materiales de Arag\'on (INMA, UNIZAR-CSIC), Zaragoza, Spain}

\author{\`Oscar Iglesias}
\email{oscariglesias@ub.edu}
\affiliation{Dpt. de Física de la Matèria Condensada, Universitat de Barcelona and IN2UB, c/ Mart\'{\i} i Franquès 1, 08028 Barcelona, Catalunya, Spain}

\author{David Serantes}
\thanks{Corresponding author: david.serantes@usc.gal}
\affiliation{Dpt. de Física Aplicada, Universidade de Santiago de Compostela, 15782 Santiago de Compostela, Galicia, Spain}
\affiliation{Instituto de Materiais (iMATUS), Universidade de Santiago de Compostela, 15782 Santiago de Compostela, Galicia, Spain}


\date{\today}

\begin{abstract}
Conventional transmission electron microscopy (TEM) records two-dimensional (2D) projections of three-dimensional (3D) objects, complicating the inference of nanoparticle morphology. This limitation is investigated for quasi-spherical iron oxide nanoparticles (IONPs) imaged after deposition on a TEM grid and after cellular internalization and immobilization in resin. A statistical inverse-reconstruction method is developed in which synthetic 3D particle populations are randomly oriented, projected onto a TEM-like plane, and compared with the experimental equivalent-diameter and axial-ratio distributions. Models of increasing geometrical complexity are considered, from monodisperse prolate spheroids to size- and shape-polydisperse prolate populations and triaxial ellipsoids. Size polydispersity is required to reproduce the projected equivalent-diameter distribution, whereas relaxing axial symmetry through a triaxial description enables simultaneous reproduction of the size and axial-ratio distributions. The inferred population represents a statistically compatible 3D morphology rather than a unique particle-by-particle reconstruction. Its physical relevance is assessed through a macrospin model incorporating the corresponding shape-anisotropy contributions. Application to magnetic fluid hyperthermia reveals a pronounced sensitivity of the predicted heating response to particle geometry and polydispersity. The approach provides a general framework for connecting conventional TEM measurements with morphology-dependent properties of nanoparticle ensembles.
\end{abstract}
\maketitle


\section{Introduction}


Transmission Electron Microscopy (TEM)~\cite{ruska1987development} enables the detailed characterization of materials at the nanometre scale~\cite{williams1996transmission} 
and has therefore become an essential tool
in materials science~\cite{spurgeon2021towards}. 
In addition to providing morphological information, TEM can be used to investigate structural and chemical properties~\cite{baig2021nanomaterials} and can be combined with several complementary techniques, including scanning transmission electron microscopy (STEM)~\cite{pennycook2011scanning}, the application of external stimuli during imaging (\textit{in situ} TEM)~\cite{sharma2024perspective}, and electron holography for mapping magnetic stray fields~\cite{midgley2009electron}.

Nanoparticles (NPs) are among the nanoscale systems that have benefited most from TEM characterization because even small variations in particle dimensions can produce substantial changes in surface-to-volume ratio and, consequently, in their physical and chemical performance~\cite{thanh2014mechanisms}.
TEM imaging has therefore become widely used to determine the size distributions of NPs with different compositions and intended applications, including catalysis~\cite{valden1998onset}, biological and chemical sensing~\cite{haes2002nanoscale}, and air and water remediation~\cite{yu2002effects,chen2007titanium}.
In this work, we focus on magnetic NPs, particularly iron oxide nanoparticles (IONPs), for which accurate determination of particle shape is as important as determination of particle size. Both parameters are major determinants of the magnetic response , although they act through partly different mechanisms.
Particle size determines the volume and magnetic moment and influences the balance between anisotropy energy and thermal fluctuations, thereby affecting magnetic relaxation and magnetization reversal.
Particle shape, in turn, determines the demagnetizing energy and gives rise to shape-anisotropy contribution~\cite{lopez2026including}.
Magnetic anisotropy is a key parameter governing the coupling between the magnetic moment to the crystallographic and morphological axes of the particle and, therefore, plays a central role in both reversal and relaxation processes. In IONPs, the effective magnetic anisotropy generally includes intrinsic magnetocrystalline and shape-anisotropy contributions, although surface, interface, and stress-induced contributions may also become relevant depending on particle size and structure.

\begin{figure}[!tbp]
    \centering
    \includegraphics[width=0.8\linewidth]{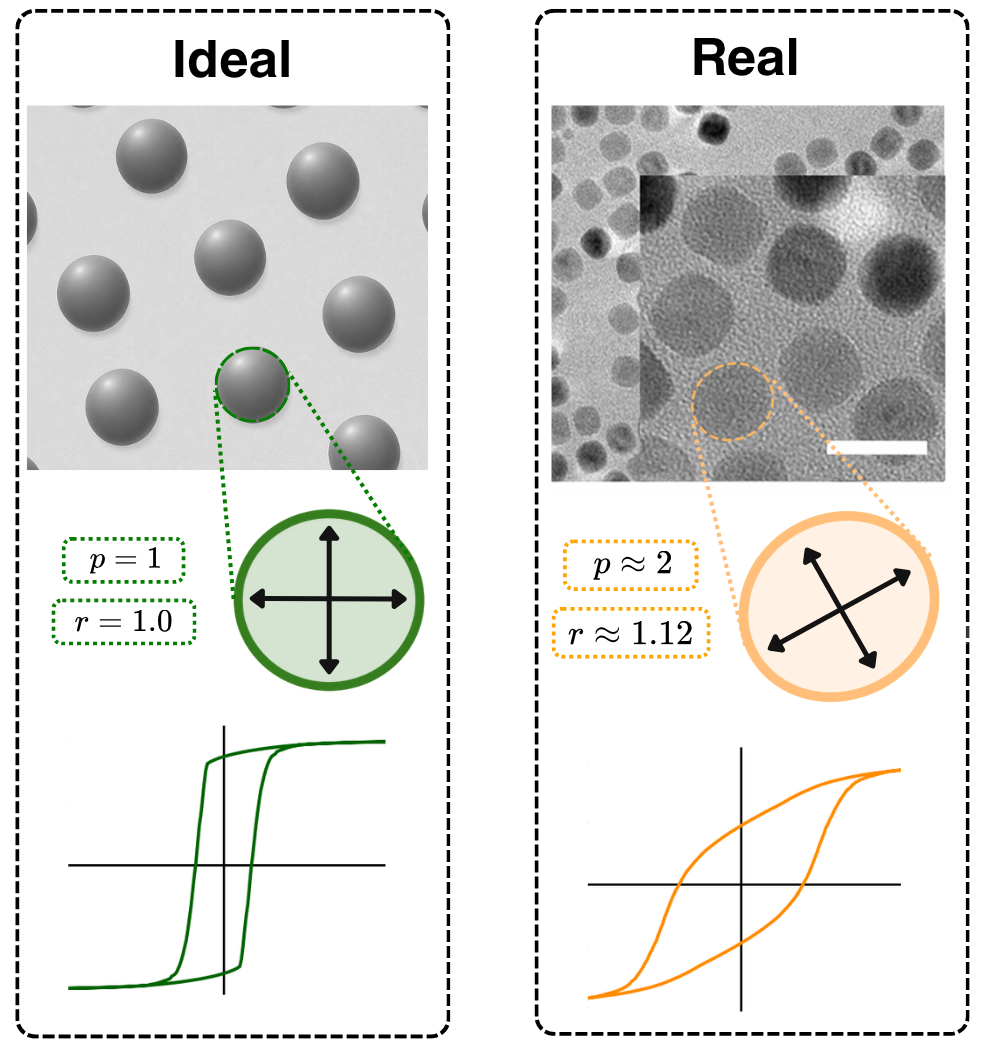}
    \caption{Schematic comparison between idealized and realistic NP morphologies, together with illustrative hysteresis loops calculated using the same material parameters, field, and temperature. The hysteresis loops are included only as qualitative examples to illustrate how changes in particle morphology can modify the magnetic response, and are not intended as a quantitative one-to-one comparison. Left: perfectly spherical particles (\(p=1\) and \(r=1\)), whose magnetic response arises solely from cubic magnetocrystalline anisotropy. Right: experimentally representative non-spherical particles, characterized by a shape exponent \(p\approx2\) and an axial ratio \(r\approx1.12\) \cite{Lopez-Vazquez_JMMM_2026}, for which the additional shape-induced anisotropy leads to a qualitatively different hysteresis response.}
    \label{fig:NPshapes}
\end{figure}

A common simplification in theoretical descriptions is to represent magnetic NPs as spheres endowed with an effective uniaxial anisotropy~\cite{conde2015single,niculaes2017asymmetric}. Although this approximation can be useful, it does not explicitly account for structural and morphological features that may strongly influence the magnetic response. 
For an ideal spherical IONP, the demagnetizing energy is isotropic and, therefore, does not give rise to shape anisotropy. Real NPs, however, are rarely perfect spheres, and even small deviations from spherical geometry can introduce a substantial shape anisotropy contribution~\cite{failde2024understanding} that can critically modify their magnetic response~\cite{lopez2026including}, as illustrated in Figure~\ref{fig:NPshapes}.
%
%
Consequently, accurate determination of both the size and geometry of IONPs is essential for understanding and predicting their magnetic properties and for optimizing their performance as contrast agents in magnetic resonance imaging (MRI)~\cite{shen2017iron}, tracers in magnetic particle imaging (MPI)~\cite{velazquez2025advances}, and heat agents for cancer treatment through magnetic fluid hyperthermia (MFH)~\cite{soetaert2020cancer}.

Detailed information about the 3D size and shape of individual IONPs can be obtained using advanced techniques such as Electron Tomography (ET)~\cite{ercius2015electron} or combined STEM imaging~\cite{vanrompay2021fast}. 
These techniques can provide valuable 3D morphological information but are comparatively demanding in terms of acquisition time, instrumentation, and data processing, which restricts the number of particles that can be analyzed routinely. 
Conventional TEM, by contrast, is particularly suitable for analyzing sufficiently large numbers of particles to obtain statistically representative distributions, especially when combined with automated image-segmentation and analysis methods that can provide size and even shape distributions. Its main limitation in this context is that a conventional TEM micrograph records only a two-dimensional (2D) projection of a 3D object.


The aim of this work is to investigate the relationship between particle sizes and shapes measured from conventional TEM images, which provide the morphological information commonly used as input for theoretical models, and the corresponding 3D particle characteristics.
Because real particles generally deviate to some extent from an ideal spherical geometry, we examine how the equivalent diameter and axial ratio $r$ (ratio of major/minor axis) determined from a projected TEM image may differ from the actual 3D descriptors of the particles. 

\begin{figure}[!tbp]
    \centering
    \includegraphics[width=0.9\linewidth]{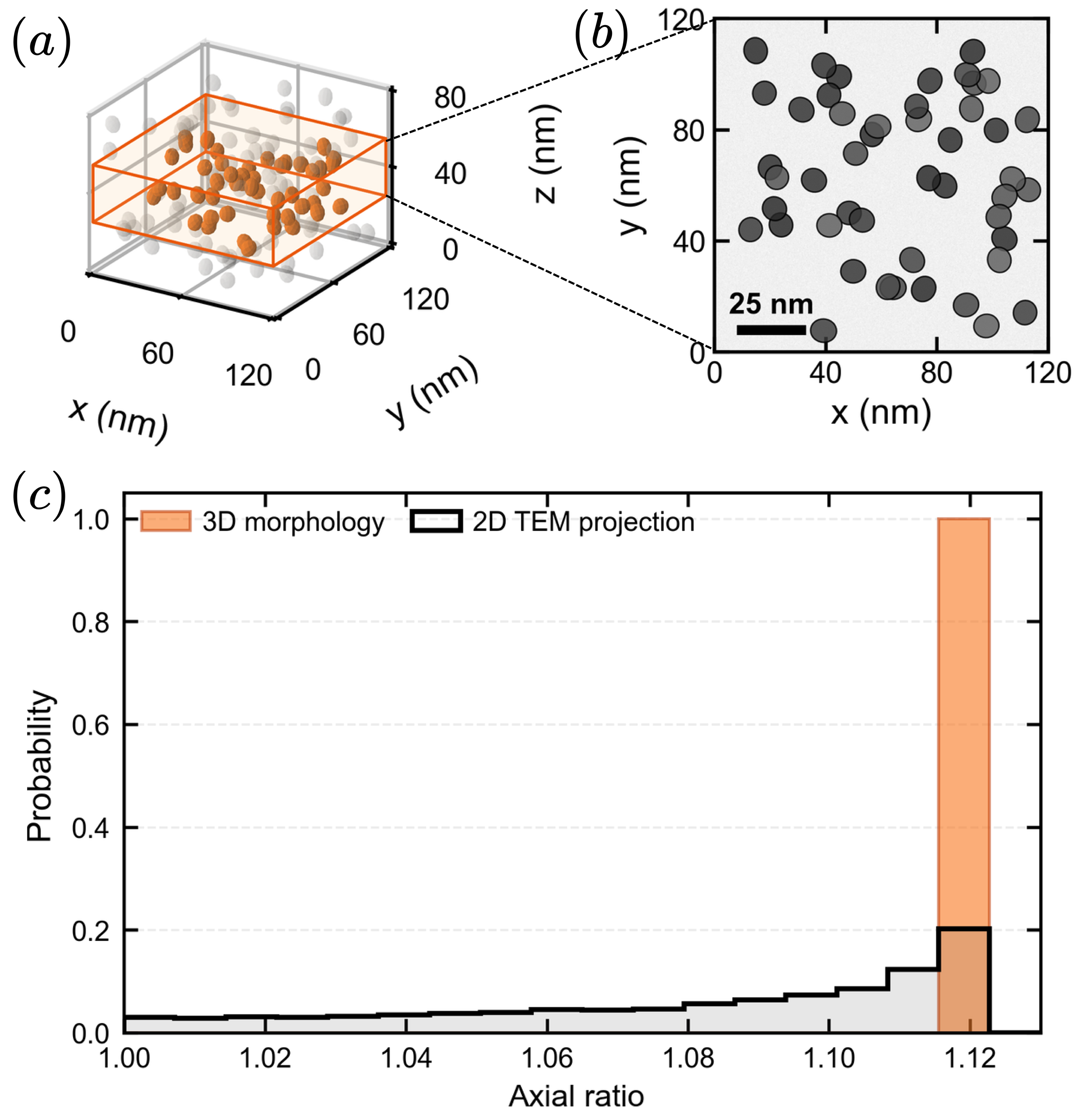}
    \caption{Illustration of the ambiguity introduced by 2D projection. (a) Example of a 3D population of randomly oriented NPs with identical size and axial ratio $r=1.12$. (b) Corresponding 2D TEM-like projection. (c) Comparison between the real 3D and the apparent 2D distribution of axial ratios obtained from the projections, highlighting the broadening produced solely by differences in particle orientation.}
    \label{fig:scheme}
\end{figure}

This projection effect is illustrated in Figure~\ref{fig:scheme} for a population of identical particles with $r = 1.12$ and uniformly random orientations.

Although the particles are monodisperse in both size and shape in 3D, their 2D projections as obtained from TEM exhibit distributions of apparent equivalent diameters and axial ratios.
Thus, an intrinsically monodisperse population of nonspherical particles can appear polydisperse when characterized using conventional TEM projections.
Because particle morphology can strongly influence the magnetic response~\cite{failde2024understanding, gavilan2021size}, it is crucial to establish how the 2D distributions obtained experimentally by TEM are related to the actual underlying 3D distributions of the particle ensemble.

Projection-induced ambiguity in NP characterization has been recognized in previous studies. Conventional TEM provides only a 2D projection of the underlying 3D structure, potentially leading to biased or non-unique estimates of particle size and shape~\cite{ercius2015electron,vanhecke2020characterization}. In particular, Vanhecke \textit{et al.} showed that preferential orientations of anisotropic IONPs on a TEM support can partially conceal their out-of-plane anisotropy. Here, we address this limitation from a statistical perspective by constructing computational models of 3D particle populations and determining which of them are compatible with the experimentally measured 2D TEM distributions.

The article is organized as follows. We first analyze the projected morphology of IONPs from a single batch under two experimental conditions: as-synthesized particles deposited from suspension onto a conventional TEM grid, and the same particles observed after cellular internalization and immobilization in a solid resin. We then develop a statistical inverse-reconstruction method to identify 3D particle populations compatible with the experimentally measured 2D equivalent-diameter and axial-ratio distributions. To this end, geometrical models of increasing complexity, ranging from prolate spheroids to polydisperse triaxial ellipsoids, are considered. Finally, the implications of the reconstructed particle morphology for magnetic fluid hyperthermia are investigated using an effective macrospin model that incorporates the corresponding shape-anisotropy contributions.

\section{Experimental scenario: analyzing IONPs size and geometry from TEM images}

IONPs with an average size of approximately 13 nm, previously described in Ref.~\cite{fernandez2024reversible}, were analyzed under two experimental conditions: as-synthesized NPs deposited from suspension onto TEM grids, and the same NPs observed after cellular internalization and embedding in a solid resin. Details of NP synthesis and characterization, as well as the cellular internalization and TEM sample-preparation procedures, are provided in Sections~\ref{sec:synthesis} and~\ref{sec:invitro}, respectively. The two datasets were analyzed using the same image-analysis procedure, described in detail in Section~\ref{sec:TEM_analysis}. 

\begin{figure*}[!tbp]
    \centering
    \includegraphics[width=1.0\textwidth]{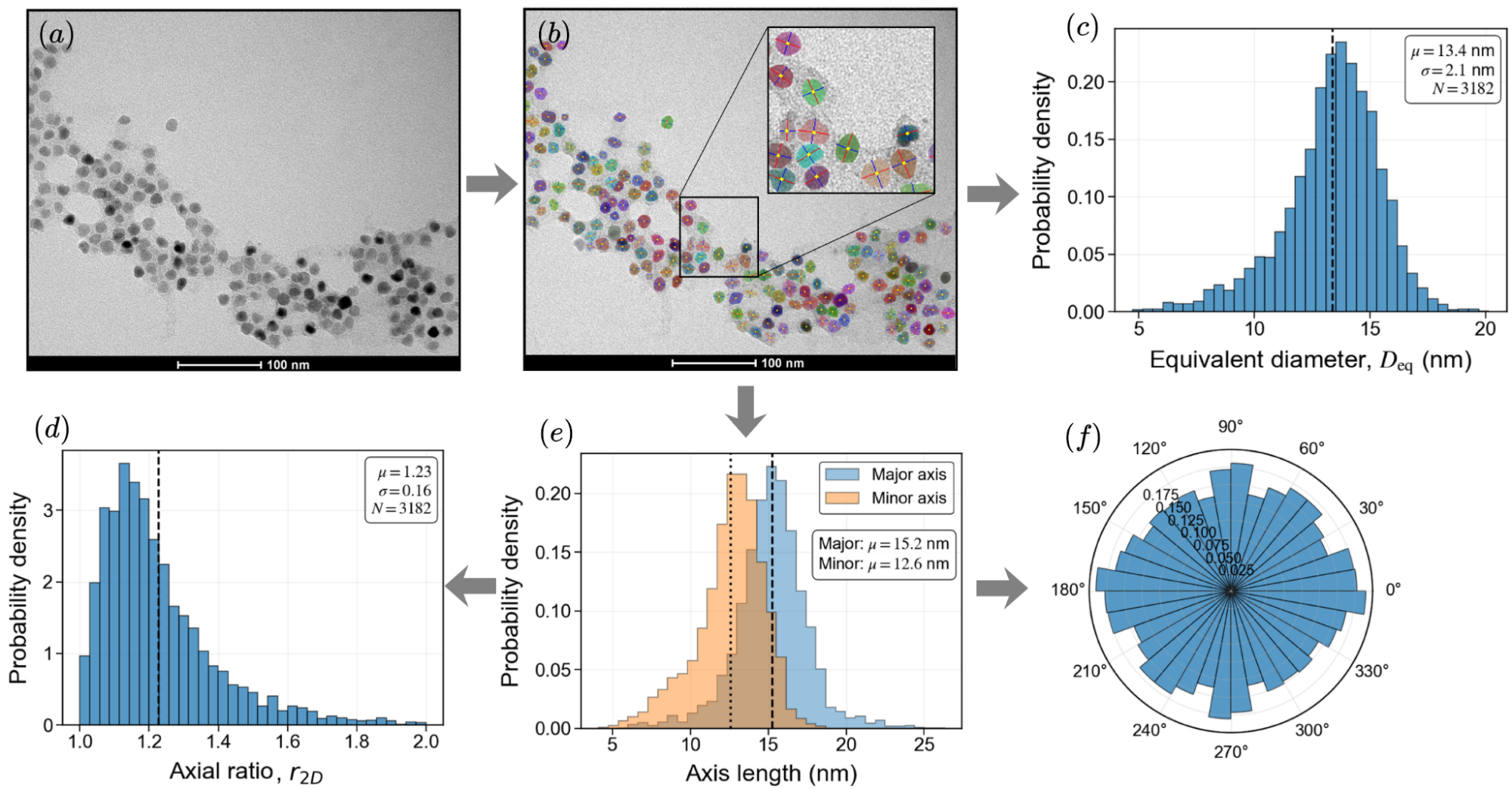}
    \caption{
TEM image analysis and projected morphological characterization of
as-synthesized IONPs deposited from suspension onto a TEM grid. (a) Representative TEM micrograph of the NP population; scale bar: 100 nm. (b) Same micrograph after CellposeSAM segmentation and PCA-based morphological analysis, with the projected major and minor axes overlaid on each detected particle. The inset shows an enlarged region illustrating the segmentation and axis determination; scale bar: 100 nm. (c) Probability-density distribution of the projected equivalent diameter, $D_{\mathrm{eq}}$.
(d) Probability-density distribution of the projected axial ratio, $r_{2D}=a_{\mathrm{major}}/a_{\mathrm{minor}}$. (e) Probability-density distributions of the projected major- and minor-axis lengths. Vertical lines indicate the corresponding mean values. (f) Polar distribution of the projected major-axis orientations in the TEM image plane. }
    \label{fig:TEM_exp-ana}
\end{figure*}

Figure~\ref{fig:TEM_exp-ana} summarizes the complete analysis for the as-synthesized NPs deposited from suspension onto a TEM grid, panel ~\ref{fig:TEM_exp-ana}(a). Briefly, individual NPs were segmented using CellposeSAM \cite{pachitariu2025cellposesam} and their projected morphology was characterized from the resulting masks using a principal component analysis (PCA)-based procedure, as shown in panel ~\ref{fig:TEM_exp-ana}(b). For each particle, the projected equivalent diameter ($D_{eq}$) and major- and minor-axis lengths were determined, panels ~\ref{fig:TEM_exp-ana}(c) and ~\ref{fig:TEM_exp-ana}(e), respectively. From the major- and minor-axes, the axial ratio ($r_{2D}$) and spatial orientation of the particles are also obtained, see panels ~\ref{fig:TEM_exp-ana}(d) and ~\ref{fig:TEM_exp-ana}(f), respectively. The corresponding analysis of the NPs observed after cellular internalization and resin embedding is provided in the Supporting Information.

The results in Figure~\ref{fig:TEM_exp-ana} show a size distribution centered at $D_{\mathrm{eq}}=13.4\pm2.1$ nm, an axial-ratio distribution of mean value $<r_{2D}>=1.23$, with a standard deviation of $\sigma_{r_{2D}}=0.16$. The separated corresponding distributions of the major- and minor-axis lengths have mean values of $15.2$ and $12.6$ nm, respectively. The polar distribution of the projected major-axis orientations does not reveal a clear preferential direction. Thus, although the individual particles display a measurable projected elongation, this elongation is not preferentially aligned along a particular in-plane direction on the TEM grid.

The same image-analysis procedure was also applied to NPs observed after cellular internalization and embedding in resin. The complete analysis is reported in the Supporting Information (Figure S1). Despite the substantially different sample environment and preparation procedure, the projected morphological characteristics remain very similar to those obtained for the suspension sample. In particular, the projected axial-ratio distribution is essentially unchanged, with a mean value of  $<r_{2D}>=1.23$ and a standard deviation of $\sigma_{r_{2D}}=0.16$. The equivalent-diameter distribution is slightly shifted towards smaller values, $D_{\mathrm{eq}}=12.9\pm1.9$ nm, compared with $13.4\pm2.1$ nm for the suspension sample, while remaining within a similar characteristic size range. The reduction in equivalent diameter may be related to intracellular degradation \cite{fernandez2022iron, lartigue2013biodegradation, levy2010degradability}. It could also be a result of the lower imaging contrast in the NP images from cell samples \cite{fernandez2022influence}. The projected major-axis orientations of the intracellular NPs also show no clear preferential direction in the TEM image plane, in agreement with the behavior observed for the suspension sample. 

Taken together, these results already reveal a slight deviation from the sphericity in the particles, from the ideal spherical-shape assumption. However, all the above quantities are descriptors of 2D projections and cannot, by themselves, unambiguously establish the underlying 3D particle morphology. This motivates the inverse-reconstruction described in the following section, in which synthetic 3D particle populations are randomly oriented and projected onto a 2D TEM-like plane to determine which NP morphologies are compatible with the experimentally observed distributions.

\section{3D reconstruction by inverse modeling procedure}

The apparent axial ratio measured in TEM depends not only on the real particle shape, but also on its orientation with respect to the electron beam. Therefore, the projected geometrical descriptors, such as the axial ratio and the equivalent diameter, obtained from 2D TEM images may not necessarily correspond directly to the real 3D morphology of the NPs. For this reason, the aim of this section is to determine which 3D particle populations could give rise to the experimental TEM distributions after random projection. 

For this purpose, we use a statistical inverse approach. Synthetic populations of 3D particles are generated, randomly oriented with respect to the TEM plane, projected onto a TEM-like plane, and compared with the experimental distributions of projected axial ratio and equivalent diameter. Several models are considered with increasing geometrical complexity. The first model assumes that all particles have the same 3D size and shape, so that the observed variability arises only from projection effects. The second model assumes polydisperse 3D sizes while keeping a monodisperse shape. The third model keeps the particle size fixed, while the 3D axial ratio is allowed to be polydisperse. The fourth model allows both the 3D equivalent diameter and the 3D axial ratio to vary. Finally, the axial symmetry imposed by the prolate spheroidal approximation is relaxed by considering triaxial ellipsoids with three different characteristic axes.

\subsection{Inverse modeling procedure}

We first define the projection model used for the prolate spheroidal cases. Each particle is approximated as a prolate spheroid described by a 3D equivalent diameter \(D_{\mathrm{eq},3D}\) and a 3D axial ratio \(r_{3D}\). The corresponding minor and major axes are
\begin{equation}
D_{\mathrm{minor},3D}
=
\frac{D_{\mathrm{eq},3D}}{r_{3D}^{1/3}}, \quad \text{and} \quad D_{\mathrm{major},3D}
=
D_{\mathrm{eq},3D} r_{3D}^{2/3}.
\end{equation}
For each simulated particle, a random orientation is assigned by sampling \(\cos\theta\sim U(0,1)\), where \(\theta\) is the angle between the spheroid major axis and the electron beam direction. The projected axial ratio is then calculated as
\begin{equation}
r_{\mathrm{2D}}
=
\sqrt{
1+
\left(
r_{3D}^{2}-1
\right)
\sin^{2}\theta
}.
\label{eq:r2d_projection}
\end{equation}
A geometrical derivation of Eq.~\eqref{eq:r2d_projection} is provided in Section~\ref{app:r2d_derivation}. This expression satisfies the expected limits: when \(\theta=0\), the particle is viewed along its major axis and the projected shape is circular, so \(r_{\mathrm{2D}}=1\); when \(\theta=90^\circ\), the full elongation is visible in the projection plane and \(r_{\mathrm{2D}}=r_{3D}\). The projected equivalent diameter is then calculated as
\begin{equation}
D_{\mathrm{eq},2D}
=
D_{\mathrm{eq},3D}
\,r_{\mathrm{2D}}^{1/2}
\,r_{\mathrm{3D}}^{-1/3}.
\end{equation}
For each prolate spheroidal model, a synthetic population of particles is generated by assigning values of \(D_{\mathrm{eq},3D}\) and \(r_{3D}\) according to the assumptions of the model. Each particle is then given a random orientation with respect to the electron beam and projected onto the TEM plane. This produces a simulated set of projected axial ratios, \(r_{\mathrm{2D,sim}}\), and projected equivalent diameters, \(D_{\mathrm{eq,sim}}\), which can be directly compared with the experimental TEM measurements.

The comparison is performed at the level of the full distributions rather than only comparing their mean values. This is important because two distributions may have similar averages but different widths, asymmetries or tails. To quantify the agreement between the experimental and simulated distributions, the one-dimensional Wasserstein distance is used \cite{panaretos2019statistical, rubner2000earth, villani2009optimal}. This metric, also known as the Earth Mover's Distance, can be intuitively understood by imagining each probability distribution as a pile of earth spread along the variable axis. The Wasserstein distance then represents the minimum cost required to move and reshape one pile until it matches the other. In the present context, this means estimating how much probability mass must be displaced along the axial-ratio or size axis to transform the simulated distribution into the experimental TEM distribution. 

The total loss function is defined as

\begin{equation} \label{eq:loss}
\mathcal{L}
=
\frac{
W(r_{\mathrm{2D,obs}},r_{\mathrm{2D,sim}})
}{
\sigma(r_{\mathrm{2D,obs}})
}
+
\frac{
W(D_{\mathrm{eq,obs}},D_{\mathrm{eq,sim}})
}{
\sigma(D_{\mathrm{eq,obs}})
}.
\end{equation}

Here, \(W\) denotes the Wasserstein distance, \(r_{\mathrm{2D,obs}}\) and \(D_{\mathrm{eq,obs}}\) are the axial ratios and equivalent diameters measured from the TEM images, and \(r_{\mathrm{2D,sim}}\) and \(D_{\mathrm{eq,sim}}\) are the corresponding quantities obtained after projecting the simulated 3D particles. Each term is normalized by the standard deviation of the corresponding experimental distribution. This normalization is introduced because the axial ratio is dimensionless and typically varies over a narrow range, whereas the equivalent diameter is measured in nanometers and may have a larger numerical spread. In this way, both the shape distribution and the size distribution contribute on comparable scales to the total loss.

The optimal 3D parameters are those that minimize \(\mathcal{L}\). A small value of the first term indicates good agreement in the projected axial-ratio distribution, whereas a small value of the second term indicates good agreement in the projected size distribution. Further details of the optimization algorithm and its numerical implementation are provided in Section~\ref{sec:opt_proc}. This procedure is first applied to prolate spheroidal models with increasing levels of polydispersity and is subsequently extended to triaxial ellipsoids. In the triaxial case, the projected axial ratio cannot be obtained analytically from Eq.~\eqref{eq:r2d_projection}, since the particles have three independent axes. Instead, it is determined directly from the full 2D projection of each randomly oriented triaxial ellipsoid as the ratio between the major and minor axes of the resulting projected ellipse.

\begin{figure*}[!tbp]
    \centering
    \includegraphics[width=1.0\textwidth]{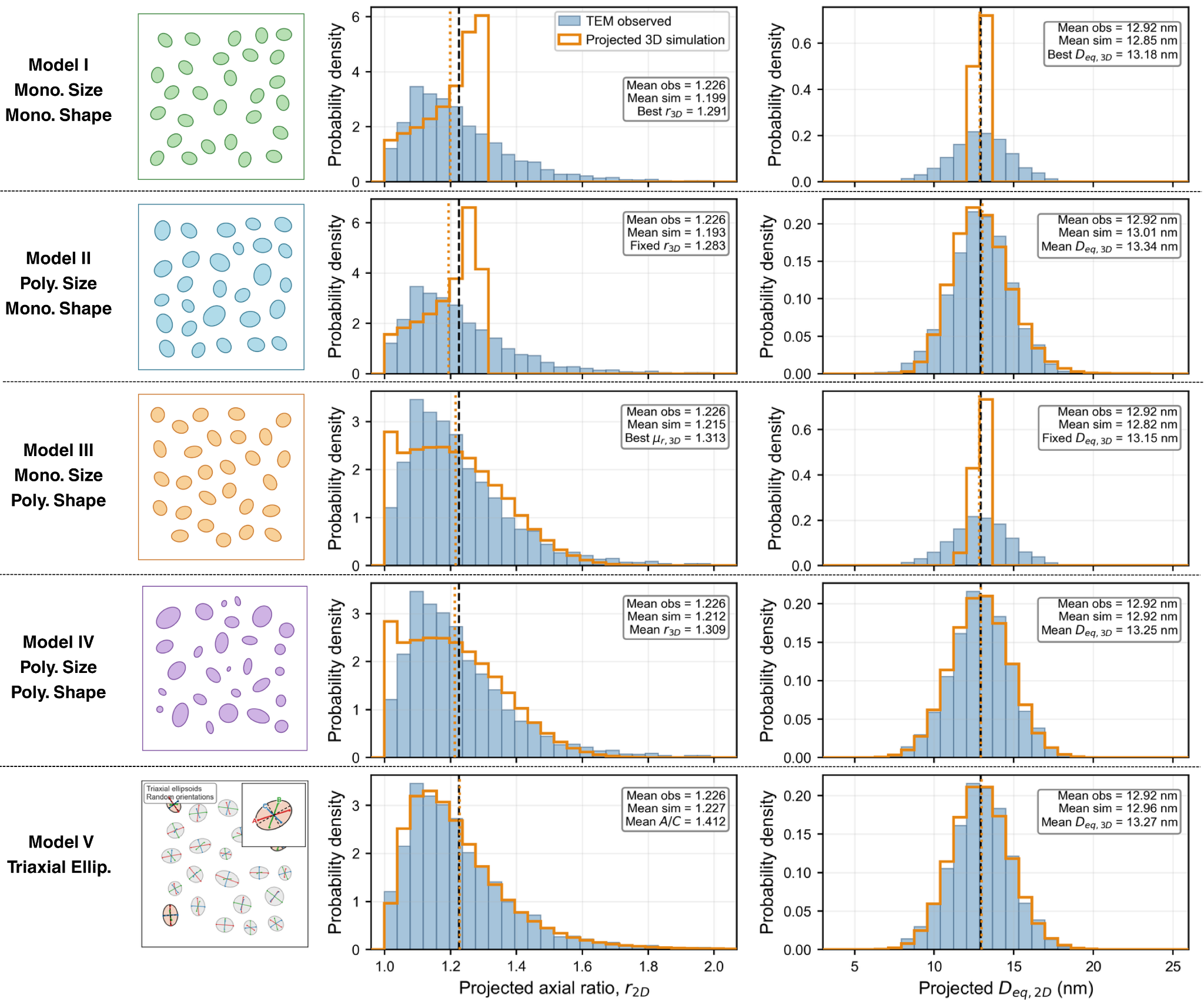}
    \caption{
Comparison of the five inverse-3D reconstruction models. From top to
bottom, the rows correspond to Model I, both size- and shape- monodisperse; Model II, size-polydisperse and shape-monodisperse; Model III, size-monodisperse and shape-polydisperse; Model IV, both size- and shape-polydisperse; and Model V, based on randomly oriented size- and shape-polydisperse triaxial
ellipsoids. In each row, the left
panel schematically illustrates the particle morphology assumed by
the corresponding model. The central and right panels compare the
experimentally observed TEM distributions with those obtained by
projecting the reconstructed 3D populations for the axial ratio,
$r_{2D}$, and equivalent diameter, $D_{\mathrm{eq},2D}$,
respectively. Blue filled histograms represent the TEM measurements,
whereas orange step histograms represent the projected 3D
simulations. The vertical dashed, dash-dotted, and dotted lines
indicate the experimental mean, the simulated mean, and the
optimized fixed parameter or inferred mean of the corresponding 3D
quantity, respectively. In Model V, the colored axes $A$, $B$, and
$C$ illustrate the three independent principal axes of the
triaxial particles.}
\label{fig:models_3D_reconstruction}
\end{figure*}

\subsection{Model I: size- and shape-monodisperse} \label{model_I}

The first model considered is the simplest possible 3D configuration. In this case, all NPs are assumed to have the same size and elongation, as schematically illustrated in the left panel of the first row of Figure~\ref{fig:models_3D_reconstruction}. Therefore,
\begin{equation}
D_{\mathrm{eq},3D}=D_0 \quad \text{and } \quad r_{3D}=r_0.
\end{equation}
Under this assumption, the only source of variability in the projected TEM distributions is the random orientation of the particles with respect to the electron beam.

The two free parameters of the model, \(D_0\) and \(r_0\), were optimized by minimizing the loss function defined above. During the optimization, the normalized Wasserstein loss progressively decreased until reaching an approximately constant value, indicating convergence towards the best solution allowed by this restricted model. After convergence, the optimized 3D geometry was projected and compared with the experimental TEM distributions. 
As shown in the first row of Figure~\ref{fig:models_3D_reconstruction}, this model is not able to reproduce the experimental distributions satisfactorily. Although the optimized values provide the best possible agreement under the assumption of identical particles, the simulated projected distributions remain considerably narrower and more constrained than the TEM data. This result indicates that the variability observed in the TEM measurements cannot be explained by random projection effects alone under the simplest assumption of size and shape monodispersity.

\subsection{Model II: size-polydisperse, shape-monodisperse}
\label{model_II}

The second model introduces size polydispersity while retaining the same axial ratio for all particles, as schematically illustrated in the left panel of the second row of Figure~\ref{fig:models_3D_reconstruction}. Therefore,

\begin{equation}
D_{\mathrm{eq},3D}\sim(\mu_D,\sigma_D)
\quad \text{while} \quad
r_{3D}=r_0.
\end{equation}

The fitted parameters are therefore $\mu_D$, $\sigma_D$, and $r_0$. This represents a natural extension of Model I, since experimentally synthesized NP populations generally exhibit a distribution of particle sizes. As shown in the second row of Figure~\ref{fig:models_3D_reconstruction}, introducing size polydispersity substantially improves the agreement with the experimental size distribution. However, the projected axial-ratio distribution remains considerably different. This result shows that size polydispersity can account for the observed variability in particle size but cannot, by itself, reproduce the experimental shape distribution. 

\subsection{Model III: size-monodisperse, shape-polydisperse}
\label{model_III}

In this model size is assumed to be the same for all particles, whereas the axial ratio is allowed to vary from particle to particle, as schematically illustrated in the left panel of the third row of Figure~\ref{fig:models_3D_reconstruction}. Therefore,

\begin{equation}
D_{\mathrm{eq},3D}=D_0
\quad \text{while} \quad
r_{3D}\sim(\mu_r,\sigma_r).
\end{equation}

The fitted parameters are now $D_0$, $\mu_r$, and $\sigma_r$, which were determined by minimizing the same normalized Wasserstein loss used for the previous models. As shown in the third row of Figure~\ref{fig:models_3D_reconstruction}, shape-polydispersity improves the description of the experimental projected axial-ratio distribution. However, since the system is now size-monodisperse, the projected equivalent-diameter distribution remains considerably more constrained than the experimental distribution. Shape polydispersity alone is therefore insufficient to reproduce simultaneously the experimentally observed size and shape distributions.

\subsection{Model IV: size- and shape-polydisperse} \label{model_IV}

In the fourth model, both size and shape are allowed to vary from particle to particle, as schematically illustrated in the left panel of the fourth row of Figure~\ref{fig:models_3D_reconstruction}. Therefore,
\begin{equation}
D_{\mathrm{eq},3D} \sim (\mu_D,\sigma_D) \quad \text{ and } \quad r_{3D} \sim (\mu_r,\sigma_r).  
\end{equation}
The fitted parameters are in this case \(\mu_D\), \(\sigma_D\), \(\mu_r\), and \(\sigma_r\). As shown in the fourth row of Figure~\ref{fig:models_3D_reconstruction}, including size polydispersity allows the projected equivalent-diameter distribution to be reproduced with high accuracy. In contrast, the agreement in the projected axial-ratio distribution does not improve substantially compared with Model III.

These results indicate that size polydispersity is essential to reproduce the experimental equivalent-diameter distribution observed in TEM. However, even when both size and shape polydispersity are included, the projected axial-ratio distribution cannot be matched with the same level of precision. This suggests that the remaining discrepancy in shape may be related to additional effects not included in the present model, such as deviations from an ideal prolate spheroidal geometry, or correlations between size and shape, for example.

\subsection{Model V: triaxial ellipsoids}\label{model_V}

Given that the analysis of our experimental data did not show any correlation between particle size and elongation, as a further step we considered a more general particle morphology, namely a tri-axial ellipsoid. In this model, each particle is represented by three different axes,
\begin{equation}
A \neq B \neq C,
\end{equation}
instead of assuming two identical minor axes as in the prolate spheroidal case. The shape is therefore described by two independent ratios, \(A/B\) and \(B/C\), together with a polydisperse equivalent diameter. For each candidate population, the triaxial ellipsoids were randomly oriented in 3D and projected onto the TEM plane, in the same way as with the previous model cases, as schematically illustrated in the left panel of the fifth row of Figure~\ref{fig:models_3D_reconstruction}. The projected axial ratio and equivalent diameter were then calculated from the resulting 2D ellipse and compared with the experimental TEM distributions using the same Wasserstein-based loss function.

As shown in the fifth row of Figure~\ref{fig:models_3D_reconstruction}, contrary to models I-IV, the triaxial model provides an accurate simultaneous reproduction of both the projected axial-ratio distribution and the projected equivalent-diameter distribution. The improved agreement indicates that relaxing the axial-symmetry constraint of the prolate models leads to a more flexible and realistic description of the experimental TEM data.

Experimentally, this description is also reasonable: allowing three independent particle axes makes it possible to account for deviations from ideal prolate spheroids, such as irregularities, faceting, or unequal transverse dimensions, which are commonly expected in real NP populations~\cite{salazar2008cubic,roca2019design}. Such deviations have been reported in particles produced via thermal decomposition, as in this particular example, because the heating rate, the precursors, co-surfactant ratios and other subtle variations in reaction conditions govern the final shape of the particles \cite{llacer2024simple, cotin2018unravelling}. 

Nevertheless, the result of Model V should not be interpreted as a unique reconstruction of the true 3D particle morphology. The inverse problem is intrinsically non-unique, and a sufficiently flexible prolate model might also be able to reproduce part of the projected TEM distributions. Therefore, the better performance of the triaxial model should be understood as evidence that the TEM-derived size and shape distributions are more naturally described when the assumption of axial symmetry of ideal ellipsoidal shape is relaxed, rather than as a definitive proof that every particle is strictly triaxial. It is also worth mentioning here that the ellipsoidal shape was chosen based on its general adequacy to describe small axial ratio particles of several shapes, ranging from pure spheres to cubes, through the general \textit{superball} geometry~\cite{lopez2026including,Lopez-Vazquez_JMMM_2026}.

\section{Application example: Magnetic fluid hyperthermia (MFH)}

To assess the practical consequences of the different 3D reconstructions we perform a theoretical study of MFH performance for the different model conditions. The magnetic response was simulated using the effective macrospin framework described in Section~\ref{sec:magnetic_model}. This reduced description was previously validated against full micromagnetic calculations for a representative triaxial NP, showing very good agreement under the conditions considered.

We therefore investigated how the small variations in particle size and shape observed in the TEM-based analysis influence the magnetic heating efficiency. In particular, we focused on the energy dissipated over the hysteresis cycle, quantified through the specific loss power (SLP). For this purpose, we simulated an ensemble of \(5000\) non-interacting ferromagnetic NPs, each modeled as a macrospin. The anisotropy axes were randomly oriented in order to represent an isotropic particle ensemble. For each NP, the magnetization dynamics was simulated over \(10\) hysteresis cycles using the stochastic Landau--Lifshitz--Gilbert equation at \(T=300~\mathrm{K}\).

The simulated particle populations were generated using the reconstructed size distribution obtained from the TEM analysis, centered around approximately \(d \approx 13\) nm. Ensembles corresponding to the five reconstruction models introduced in Sections~\ref{model_I}--\ref{model_V} were generated using their respective 3D morphological parameters. Models I--IV were represented as prolate spheroids, with their shape incorporated through the corresponding effective uniaxial shape
anisotropy, whereas Model V was described using the reconstructed triaxial geometry and the associated biaxial shape-anisotropy contribution detailed in Section~\ref{sec:magnetic_model}. In all cases, the cubic magnetocrystalline anisotropy of magnetite was included using \(K_{\mathrm{cub}}=-1.1\times10^{4}~\mathrm{J\,m^{-3}}\), with randomly oriented cubic anisotropy axes.

Simulations were performed at two excitation frequencies, \(f=100~\mathrm{kHz}\) and \(f=1~\mathrm{MHz}\), under the same AC magnetic-field amplitude. In the main text, we focus on the results obtained at \(100~\mathrm{kHz}\), while the corresponding results at \(1~\mathrm{MHz}\) are provided in the Supporting Information (Figure S6).

For each case, the SLP was calculated as
\begin{equation}
SLP= \frac{E_{\mathrm{loss}} f}{\rho},
\end{equation}
where \(E_{\mathrm{loss}}\) is the energy dissipated per cycle, \(f\) is the field frequency and \(\rho\) is the material density.

\begin{figure}[!tbp]
    \centering
    \includegraphics[width=1.0\linewidth]{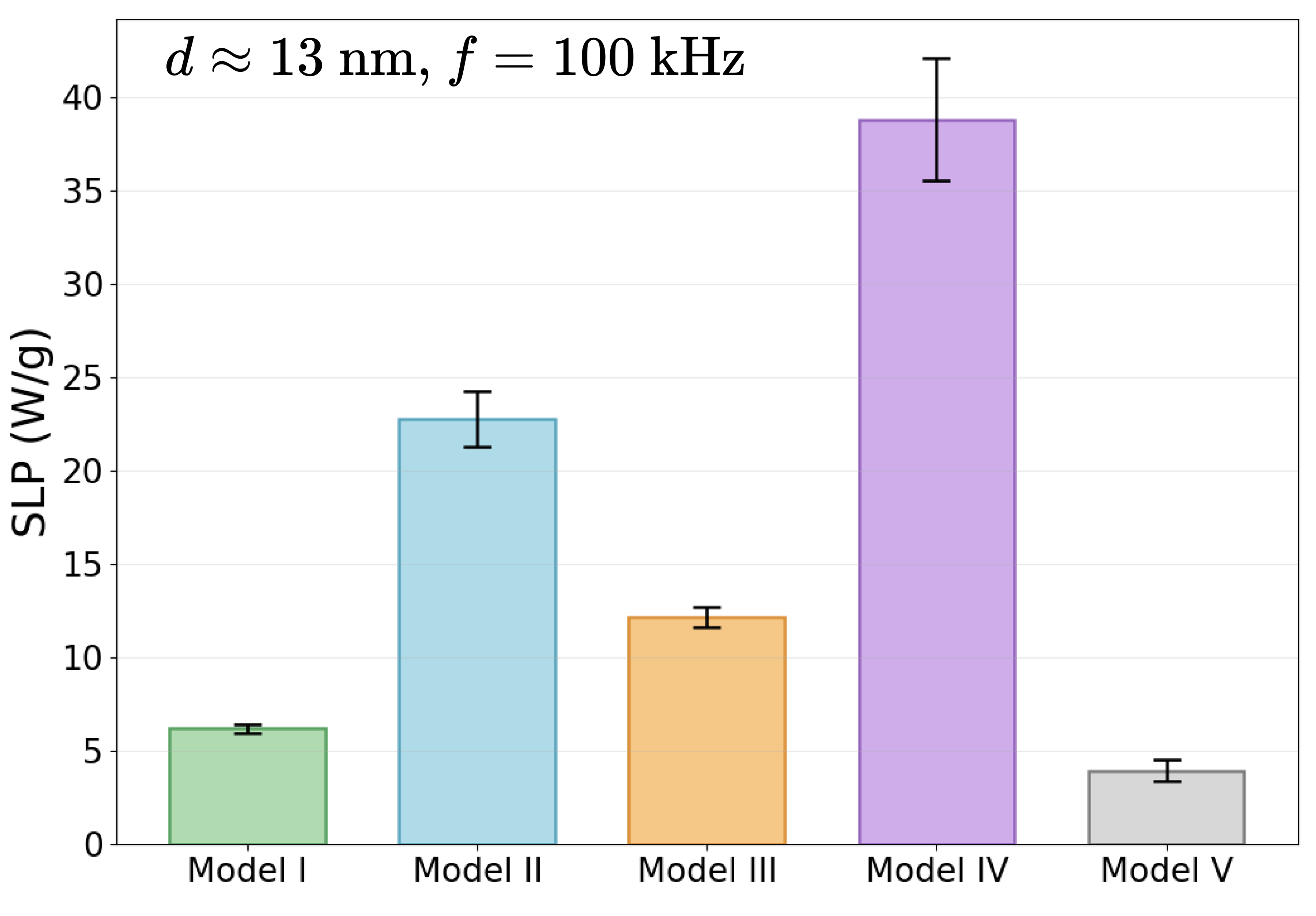}
    \caption{
    Specific loss power (SLP) predicted for NP ensembles generated according to the five 3D morphological models obtained from the inverse TEM reconstruction described in Sections~\ref{model_I}--\ref{model_V}. The particle-size distributions correspond to the reconstructed population centered around approximately \(d \approx 13\) nm. Results were obtained at \(f=100~\mathrm{kHz}\) and \(\mu_0H_{\max}=50~\mathrm{mT}\). Models I--IV correspond to prolate descriptions with different assumptions regarding size and shape polydispersity, whereas Model V represents a polydisperse population of triaxial ellipsoids. Error bars indicate the standard deviation obtained from the simulated ensemble.
    }
    \label{fig:SLP_5_models}
\end{figure}

Figure~\ref{fig:SLP_5_models} summarizes the SLP predicted at \(f=100~\mathrm{kHz}\) for the reconstructed NP population centered around \(d\approx13\) nm. The predicted SLP depends strongly on the geometrical description. Models I--IV span a broad range of heating efficiencies, with the lowest SLP obtained for Model I and the highest for Model IV. Importantly, Models II and III yield clearly different values, showing that introducing size polydispersity and shape polydispersity individually does not have the same effect on the predicted magnetic losses. In contrast, Model V, based on triaxial particles and biaxial shape anisotropy, yields a substantially smaller SLP than the most strongly heating prolate descriptions.

These results demonstrate that the predicted heating efficiency is highly sensitive to the 3D morphological description inferred from TEM. In particular, the differences between the prolate descriptions (Models I--IV) and the triaxial one (Model V) highlight the importance of accounting for the biaxial shape-anisotropy contribution associated with the reconstructed particle geometry. Thus, even when the morphological models are derived from the same experimental TEM dataset, different assumptions regarding the underlying 3D particle morphology can lead to substantially different predictions of magnetic heating efficiency.

The corresponding simulations performed at \(f=1~\mathrm{MHz}\), provided in the Supporting Information (Figure S6), show similar qualitative trends among the five geometrical models, although the absolute magnitude of the SLP changes with frequency.

\section{Conclusions}

In this work, we investigated how the size and shape distributions measured from conventional 2D TEM images are related to the underlying 3D morphology of nominally quasi-spherical IONPs. Analysis of particles from the same synthesis batch under two experimental conditions, deposited from suspension onto a conventional TEM grid and observed after cellular internalization and immobilization in resin, revealed deviations from a perfectly spherical geometry, with projected axial ratios centered around $r\approx 1.2$. 
Despite the different sample environments and preparation procedures, the two datasets exhibited comparable projected morphological distributions, supporting their interpretation as observations of the same underlying particle population.

To address the ambiguity introduced by 2D projection, we developed a statistical inverse-reconstruction approach in which synthetic 3D particle populations were randomly oriented, projected onto a TEM-like plane, and compared with the experimentally measured equivalent-diameter and axial-ratio distributions. Geometrical models of increasing complexity were considered, ranging from identical prolate spheroids to size- and shape-polydisperse prolate populations and, finally, to polydisperse triaxial ellipsoids. Among the models investigated, the triaxial-ellipsoid population provided the most consistent simultaneous description of the experimental equivalent-diameter and axial-ratio distributions. This result indicates that a model involving a single elongation axis is insufficient to reproduce all the morphological information contained in the TEM projections and that three distinct particle dimensions must be considered to obtain a statistically consistent representation.

The inverse projection problem is, however, intrinsically non-unique. The inferred particle population should therefore not be interpreted as a unique reconstruction of the geometry of each individual particle, but rather as a statistically compatible representation of the underlying 3D morphology. Within this limitation, the proposed approach provides substantially more realistic structural information than that obtained by directly identifying the measured 2D distributions with the corresponding 3D particle properties. It also offers a practical route for extracting population-level 3D morphological descriptors from conventional TEM datasets, for which a statistically representative number of particles can be analyzed.

The physical relevance of the reconstructed morphology was assessed by incorporating the inferred triaxial geometries into an effective macrospin model for magnetic fluid hyperthermia. The simulations revealed a pronounced sensitivity of the predicted heating response to particle size, shape, and morphological polydispersity. In particular, apparently small deviations from spherical geometry modify the shape-anisotropy contribution and can produce appreciable differences in the calculated hysteresis losses and heating performance. Consequently, models based on spherical particles or on oversimplified uniaxial geometries may fail to capture important features of the magnetic response, even for particle populations that appear approximately spherical in conventional TEM images.

Overall, our results demonstrate that particle morphology inferred directly from 2D TEM projections cannot, in general, be used as an uncorrected representation of the underlying 3D population. Although conventional TEM does not permit a unique particle-by-particle reconstruction, combining statistically representative experimental distributions with forward projection of physically plausible 3D models makes it possible to constrain the range of compatible particle geometries. The marked dependence of the simulated magnetic-fluid-hyperthermia response on these geometries emphasizes that realistic 3D structural descriptors are essential for quantitatively meaningful modeling of nominally quasi-spherical IONP systems. More broadly, the methodology developed here provides a framework for connecting routinely accessible 2D microscopy data with 3D morphology-dependent physical properties in NP ensembles.

\section{Materials and methods}

This section summarizes the experimental procedures used for NP synthesis, sample preparation, TEM imaging and image analysis, together with the geometrical framework used for the inverse-reconstruction and the magnetic modeling approach employed to describe the reconstructed particle morphologies.

\subsection{Magnetic nanoparticle synthesis and characterization} \label{sec:synthesis}

IONP synthesis was performed by thermal decomposition in organic media following already described protocols \cite{fernandez2024reversible}. Then, the oleic-acid coated hydrophobic NPs were transferred to water using an amphiphilic polymer (PMAO, poly(maleic anhydride-alt-1-octadecene)) for the conjugation with the oleic acid coating from the NPs surface \cite{fernandez2024reversible}.

Samples for the TEM observation were prepared by placing a drop of a diluted colloid suspension on a carbon coated TEM grid and allowing it to dry at room temperature overnight. Images were acquired using a Tecnai G2 TEM (FEI) operated at 200 kV.

\subsection{In vitro study and TEM sample preparation} \label{sec:invitro}

For the in vitro experiments, in order to improve the NPs internalization by the cells, the same NPs as described above were functionalized with glucose following previously described protocols \cite{fernandez2024reversible}. The NPs were sterilized with a filter of 0.22 $\mu$m. The NPs internalization was performed using a RAW 264.7 macrophage adherent mouse cell line (ATCC TIB-71) cultured as previously described \cite{fernandez2024reversible}. Cells were incubated overnight with 1 $\mu$g/mL of iron in the form of NPs and fixed with 2.5 $\%$ glutaraldehyde, postfixed with osmium, rinsed, dehydrated and embedded in resin. Ultrathin sections were prepared and mounted on TEM grids. TEM images were acquired using a Tecnai T20  (FEI Company, USA) microscope operated at 200 kV.

\subsection{TEM image analysis} \label{sec:TEM_analysis}

For the TEM image-analysis procedure, the first step consisted of identifying and segmenting the individual NPs in each image. For this purpose, the segmentation model CellposeSAM was used \cite{stringer2021cellpose, pachitariu2025cellposesam}. An approximate particle diameter (15 nm) was provided as the initial characteristic size, allowing the model to identify and segment the visible NPs. Each detected particle was assigned an individual segmentation mask. The resulting masks were visually inspected to minimize missed detections and to identify segmentation artifacts, such as merged particles or poorly resolved contours.

Once the particles had been segmented, their projected geometrical properties were extracted from the corresponding masks using a custom Python script. The particles were approximated as projected ellipsoidal objects, and their characteristic dimensions were determined using a principal component analysis (PCA)-based procedure. For each particle, the coordinates of all pixels belonging to its mask were used to determine the principal directions of the projected shape. The maximum extension of the mask along the first principal direction was assigned to the projected major axis, \(a_{\mathrm{major}}\), whereas the maximum extension along the perpendicular direction was assigned to the projected minor axis, \(a_{\mathrm{minor}}\).

The projected axial ratio was then calculated as

\begin{equation}
    r =
    \frac{a_{\mathrm{major}}}{a_{\mathrm{minor}}}.
    \label{eq:axial_ratio_main}
\end{equation}

The projected equivalent diameter was obtained from the area of each segmentation mask. These measurements were used to construct the distributions of equivalent diameter, axial ratio and major- and minor-axis lengths for both experimental conditions. Particles presenting unrealistically large axial ratios were excluded from the statistical analysis, since these values were generally associated with partially merged particles, incomplete segmentation or poorly resolved contours.

\subsection{Geometrical derivation of the projected axial ratio}
\label{app:r2d_derivation}

Here we derive the expression used to calculate the apparent axial ratio of
a prolate spheroid after projection onto the TEM image plane. The spheroid
is assumed to have one major axis \(D_{\mathrm{major},3D}\) and two equal
minor axes \(D_{\mathrm{minor},3D}\). Its real 3D axial ratio
is therefore

\begin{equation}
r_{3D}
=
\frac{
D_{\mathrm{major},3D}
}{
D_{\mathrm{minor},3D}
}.
\end{equation}

The electron beam is taken as the projection direction. Let \(\theta\) be
the angle between the spheroid major axis and the electron beam. Since the
spheroid is axisymmetric, one of the projected axes remains equal to the
minor axis of the spheroid, $D_{\mathrm{minor},2D}
=
D_{\mathrm{minor},3D}$.

The other projected axis corresponds to the maximum apparent length of the
spheroid in the projection plane. To obtain it explicitly, we consider the
2D section of the spheroid containing both the major axis and
the projection direction. This section is an ellipse with semimajor and semiminor  axes
\begin{equation}
a
=
\frac{
D_{\mathrm{major},3D}
}{2}\hspace{1cm}
b
=
\frac{
D_{\mathrm{minor},3D}
}{2}.
\end{equation}

Let \(u\) be the local coordinate along the major axis of this ellipse and
\(v\) the local coordinate along its minor axis. A point on the boundary of
the ellipse can be parametrized as $u
=
a\cos\varphi,
v
=
b\sin\varphi$,
where \(\varphi\) parametrizes the contour.

When the ellipse is projected onto the image plane,  the
projected coordinate \(X\) of a generic boundary point can be written as

\begin{equation}
X
=
u\sin\theta
+
v\cos\theta.
\end{equation}

Substituting the contour parametrization gives

\begin{equation}
X(\varphi)
=
a\sin\theta\cos\varphi
+
b\cos\theta\sin\varphi.
\end{equation}

The projected semimajor axis is not simply the projection of the 3D major
axis. Instead, it is the maximum value of \(X(\varphi)\) over all points of
the elliptical contour. Defining $P
=
a\sin\theta$ and $Q
=
b\cos\theta,$
we can write
\begin{equation}
X(\varphi)
=
P\cos\varphi
+
Q\sin\varphi.
\end{equation}

The maximum is found by differentiating with respect to \(\varphi\), giving
\begin{equation}
\tan\varphi
=
\frac{Q}{P}.
\end{equation}

At this point of the contour,
$\cos\varphi
=
\frac{P}{\sqrt{P^2+Q^2}}$ and 
$\sin\varphi
=
\frac{Q}{\sqrt{P^2+Q^2}}$.
Substituting these values into \(X(\varphi)\) it is straightforward to show that the maximum projected
coordinate 
\begin{equation}
X_{\max}
=
\sqrt{P^2+Q^2} =
\sqrt{
a^2\sin^2\theta
+
b^2\cos^2\theta
}.
\end{equation}

Since \(X_{\max}\) is the projected semimajor axis, the projected major
diameter is $D_{\mathrm{major},2D}
=
2X_{\max}$, and given that
$a
=
D_{\mathrm{major},3D}/
2$ and  $b
=
D_{\mathrm{minor},3D}/
2$, we obtain
\begin{equation}
D_{\mathrm{major},2D}^{2}
=
D_{\mathrm{major},3D}^{2}\sin^{2}\theta
+
D_{\mathrm{minor},3D}^{2}\cos^{2}\theta.
\end{equation}

The projected axial ratio is

\begin{equation}
r_{\mathrm{2D}}
=
\frac{
D_{\mathrm{major},2D}
}{
D_{\mathrm{minor},2D}
}.
\end{equation}

Using \(D_{\mathrm{minor},2D}=D_{\mathrm{minor},3D}\), we obtain

\begin{equation}
r_{\mathrm{2D}}^{2}
=
\frac{
D_{\mathrm{major},2D}^{2}
}{
D_{\mathrm{minor},3D}^{2}
}.
\end{equation}

Substituting the expression for \(D_{\mathrm{major},2D}^{2}\) gives
\begin{equation}
r_{\mathrm{2D}}^{2}
=
\left(
\frac{
D_{\mathrm{major},3D}
}{
D_{\mathrm{minor},3D}
}
\right)^2
\sin^{2}\theta
+
\cos^{2}\theta,
\end{equation}
and since
\begin{equation}
\frac{
D_{\mathrm{major},3D}
}{
D_{\mathrm{minor},3D}
}
=
r_{3D},
\end{equation}
we have
\begin{equation}
r_{\mathrm{2D}}^{2}
=
1
+
\left(
r_{3D}^{2}
-
1
\right)
\sin^{2}\theta.
\end{equation}

Therefore,

\begin{equation}
r_{\mathrm{2D}}
=
\sqrt{
1+
\left(
r_{3D}^{2}
-
1
\right)
\sin^{2}\theta
}.
\end{equation}

This expression gives \(r_{\mathrm{2D}}=1\) for \(\theta=0\), corresponding
to a spheroid viewed along its major axis, and \(r_{\mathrm{2D}}=r_{3D}\)
for \(\theta=90^\circ\), corresponding to a spheroid viewed perpendicular
to its major axis.

\subsection{Optimization procedure} \label{sec:opt_proc}

The reconstruction of the underlying 3D morphology was formulated as an inverse optimization problem. For each geometrical model, a parameter vector $\mathbf{p}$ was defined containing the unknown descriptors of the corresponding 3D
particle population. Depending on the model considered, these parameters
may describe characteristic particle dimensions, their variability, and
the parameters or statistical weights defining the underlying size and
shape distributions. The optimal parameter set was therefore defined as

\begin{equation}
\mathbf{p}^{*}
=
\operatorname*{arg\,min}_{\mathbf{p}\in\Omega}
\mathcal{L}(\mathbf{p}),
\end{equation}

where $\Omega$ denotes the prescribed parameter space and $\mathcal{L}$ is the normalized Wasserstein loss defined in Eq.~\ref{eq:loss}. For each candidate parameter vector $\mathbf{p}$, a synthetic population of $N_{\mathrm{MC,fit}}$ 3D particles was generated according to the assumptions of the corresponding model. Particle sizes and shapes were sampled from the distributions defined by $\mathbf{p}$, while their orientations were independently randomized to represent an isotropic ensemble with respect to the electron-beam direction. The resulting particles were subsequently projected onto a TEM-like plane. From these projections, the apparent axial ratio, $r_{2D,\mathrm{sim}}$, and equivalent diameter, $D_{\mathrm{eq,sim}}$, were obtained for every particle, producing simulated distributions that could be directly compared with the experimental TEM measurements. The quality of each candidate solution was then quantified through the loss function $\mathcal{L}$, which simultaneously evaluates the agreement between the experimental and simulated axial-ratio and equivalent-diameter distributions.

The minimization of $\mathcal{L}$ was performed using the differential-evolution (DE) global optimization algorithm~\cite{storn1997differential}. Unlike local optimization methods that evolve a single candidate solution, differential evolution operates with a population of parameter vectors distributed throughout the allowed search domain. At each optimization generation, new trial vectors are generated by combining and perturbing members of the current population through recombination. Each trial vector $\mathbf{u}_{i}$ is then evaluated by generating and projecting the corresponding synthetic 3D particle population and calculating its associated loss. The selection step can be written schematically as

\begin{equation}
\mathbf{p}_{i}^{(g+1)}
=
\begin{cases}
\mathbf{u}_{i},
&
\mathcal{L}(\mathbf{u}_{i})
<
\mathcal{L}\!\left(\mathbf{p}_{i}^{(g)}\right),
\\[4pt]
\mathbf{p}_{i}^{(g)},
&
\mathrm{otherwise},
\end{cases}
\end{equation}

where $g$ denotes the optimization generation. In this way, trial parameter sets that provide a better simultaneous description of the projected size and shape distributions replace poorer solutions, and the population progressively evolves towards regions of parameter space associated with lower values of $\mathcal{L}$. This population-based global optimization strategy is particularly suitable here because the objective function is obtained through Monte Carlo projection and is not available as a simple analytical function of the model parameters. The optimization was continued until the differential-evolution population converged towards a stable minimum of the objective function. The parameter vector associated with the smallest value of $\mathcal{L}$ was then identified as the reconstructed solution, $\mathbf{p}^{*}$.

\subsection{Magnetic modeling framework}
\label{sec:magnetic_model}

When modeling the magnetic response of IONPs within the single-domain regime, the macrospin approximation is commonly employed because it replaces the spatially resolved magnetization with a single magnetic moment, substantially reducing the computational cost. Here, we construct a macrospin model following the procedure introduced in our previous work \cite{lopez2026including}, incorporating the reconstructed triaxial morphology through effective anisotropy contributions. The validity of this simplified description is then assessed by comparison with full micromagnetic simulations that explicitly resolve the particle geometry and the internal magnetization distribution. This comparison allows us to determine whether the magnetic effects associated with the reconstructed morphology can be accurately captured by an effective macrospin model suitable for subsequent ensemble simulations.

\subsubsection{Micromagnetic model}

\begin{figure*}[!tbp]
    \centering
    \includegraphics[width=1.0\textwidth]{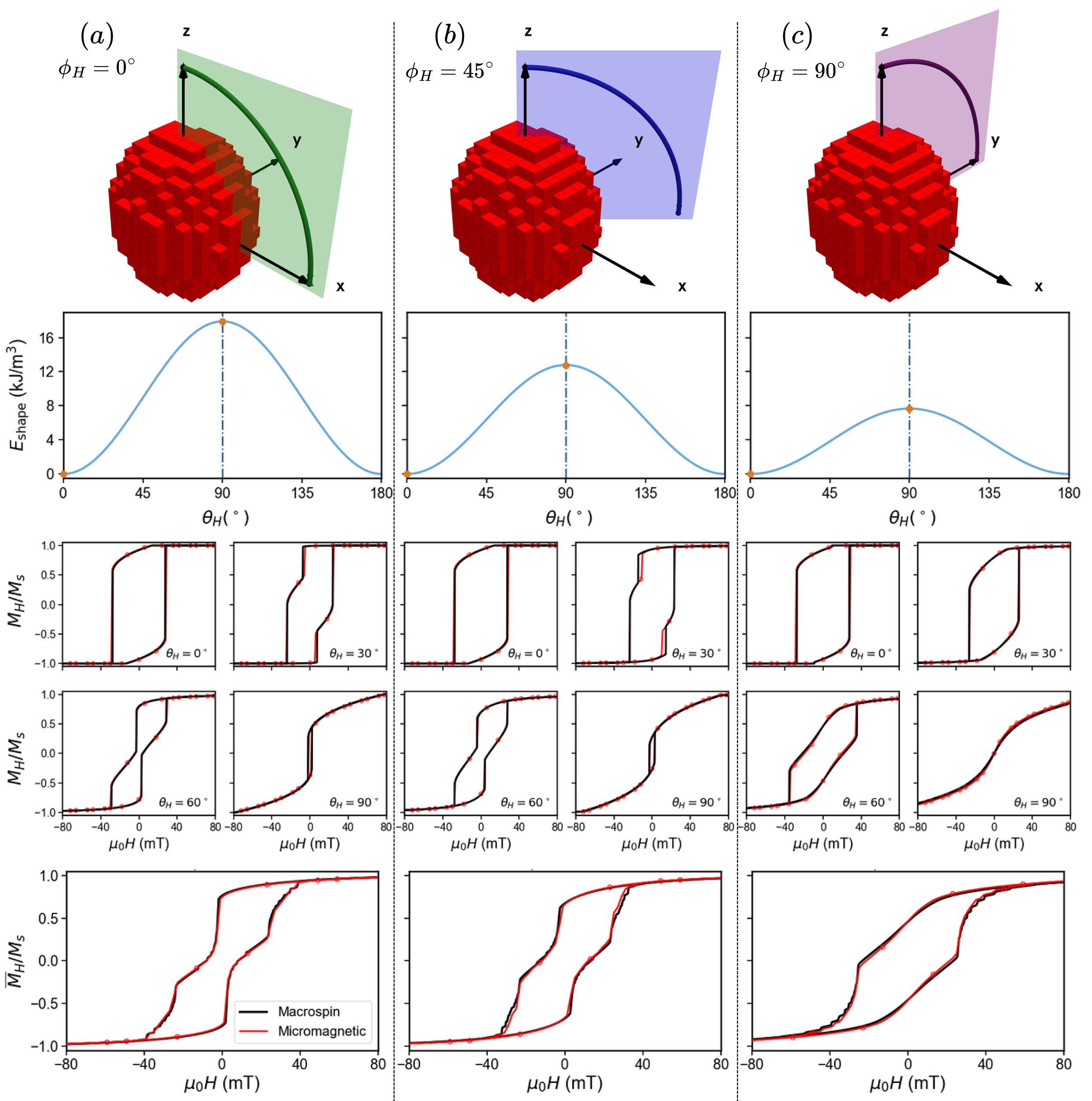}
    \caption{
Combined micromagnetic and macrospin description of the magnetic response of the representative \(15\times13\times11~\mathrm{nm}^3\) triaxial NP for three azimuthal field directions: (a) \(\phi_H=0^\circ\), (b) \(\phi_H=45^\circ\), and (c) \(\phi_H=90^\circ\). In each column, the upper panel shows the plane in which the applied field is rotated from the long particle axis (\(z\)) towards the transverse plane. For \(\phi_H=0^\circ\), the rotation occurs in the \(xz\) plane, from the long axis towards the intermediate axis, whereas for \(\phi_H=90^\circ\) it occurs in the \(yz\) plane, from the long axis towards the short axis; \(\phi_H=45^\circ\) represents an intermediate azimuthal direction. The second row shows the corresponding shape-anisotropy energy density as a function of the polar angle \(\theta_H\). The following two rows compare representative individual hysteresis loops from the macrospin (black lines) and micromagnetic (red lines and open symbols) calculations for \(\theta_H=0^\circ\), \(30^\circ\), \(60^\circ\), and \(90^\circ\). The bottom row shows the corresponding hysteresis loops after orientational averaging over \(\theta_H\).}
    \label{fig:micro_macro_combined}
\end{figure*}

The micromagnetic response of an individual NP was simulated using OOMMF \cite{donahue1999oommf}. The particle was modeled as a triaxial ellipsoid with three different characteristic axes, chosen to be representative of the average dimensions obtained from the TEM analysis. In particular, we considered a NP with axes

\begin{equation}
A = 15~\mathrm{nm}, \qquad
B = 13~\mathrm{nm}, \qquad
C = 11~\mathrm{nm}.
\end{equation}
The principal particle axes were aligned with the Cartesian coordinate system such that the longest axis \(A=15~\mathrm{nm}\) lies along \(z\), the intermediate axis \(B=13~\mathrm{nm}\) along \(x\), and the shortest axis \(C=11~\mathrm{nm}\) along \(y\). The simulations were performed using material parameters corresponding to magnetite (Fe$_3$O$_4$), summarized in Table~\ref{tab:parameters}.

\begin{table}[!tbp]
\centering
\caption{Material parameters and simulation conditions used for magnetite (Fe$_3$O$_4$).}
\label{tab:parameters}
\begin{tabular}{@{}lll@{}}
\hline
\textbf{Quantity} & \textbf{Symbol} & \textbf{Value} \\
\hline
Saturation magnetization 
& $M_s$ 
& $4.8\times10^5$ A m$^{-1}$ \\

Exchange stiffness constant 
& $A_{ex}$ 
& $1.1\times10^{-11}$ J m$^{-1}$ \\

Cubic anisotropy constant 
& $K_c$ 
& $-1.1\times10^{4}$ J m$^{-3}$ \\

Exchange length 
& $L_{\mathrm{ex}}=\sqrt{2A/(\mu_0 M_s^2)}$ 
& $8.7$ nm \\

Temperature 
& $T$ 
& $0$ K \\

Simulation approach 
& -- 
& Energy minimization \\
\hline
\end{tabular}
\end{table}

The geometry was discretized using a cubic cell size of \(1~\mathrm{nm}\). This value is well below the exchange length of magnetite (see Table \ref{tab:parameters}), ensuring a sufficiently fine spatial resolution. The total magnetic energy included the Zeeman energy, magnetocrystalline cubic anisotropy, exchange energy and demagnetizing energy,

\begin{equation}
E_{\mathrm{tot}}
=
E_{\mathrm{Z}}
+
E_{\mathrm{cub}}
+
E_{\mathrm{ex}}
+
E_{\mathrm{demag}}.
\end{equation}
A schematic representation of the simulated NP discretized on the OOMMF mesh, together with the applied-field rotation planes considered in the calculations, is shown in the top row of Figure~\ref{fig:micro_macro_combined}.

\subsubsection{Macrospin description}

In the macrospin approximation, the NP is treated as a single magnetic moment,

\begin{equation}
\boldsymbol{\mu}=M_s V \mathbf{m},
\end{equation}
where \(M_s\) is the saturation magnetization, \(V\) is the particle volume and \(\mathbf{m}\) is the reduced magnetization vector, with \(|\mathbf{m}|=1\). This approximation assumes that the magnetization remains spatially coherent during reversal.

To account for the triaxial particle geometry, an effective shape-anisotropy contribution was introduced from the demagnetizing factors of an ellipsoid with semi-axes \(a>b>c\). The corresponding demagnetizing factors, \(N_a\), \(N_b\) and \(N_c\), satisfy \(N_a+N_b+N_c=1\), with \(N_a<N_b<N_c\). Therefore, the longest axis is the easiest magnetic direction, whereas the shortest axis is the hardest one. Taking the long axis \(a\) as the reference direction, the shape-anisotropy energy density can be written as

\begin{equation}
\varepsilon_{\mathrm{shape}}
=
K_b m_b^2
+
K_c m_c^2,
\end{equation}

where \(m_b\) and \(m_c\) are the magnetization components along the intermediate and short particle axes, respectively, and

\begin{equation}
K_b=
\frac{1}{2}\mu_0 M_s^2\left(N_b-N_a\right),
\qquad
K_c=
\frac{1}{2}\mu_0 M_s^2\left(N_c-N_a\right).
\end{equation}

\begin{figure*}[!tbp]
    \centering
    \includegraphics[width=0.9\textwidth]{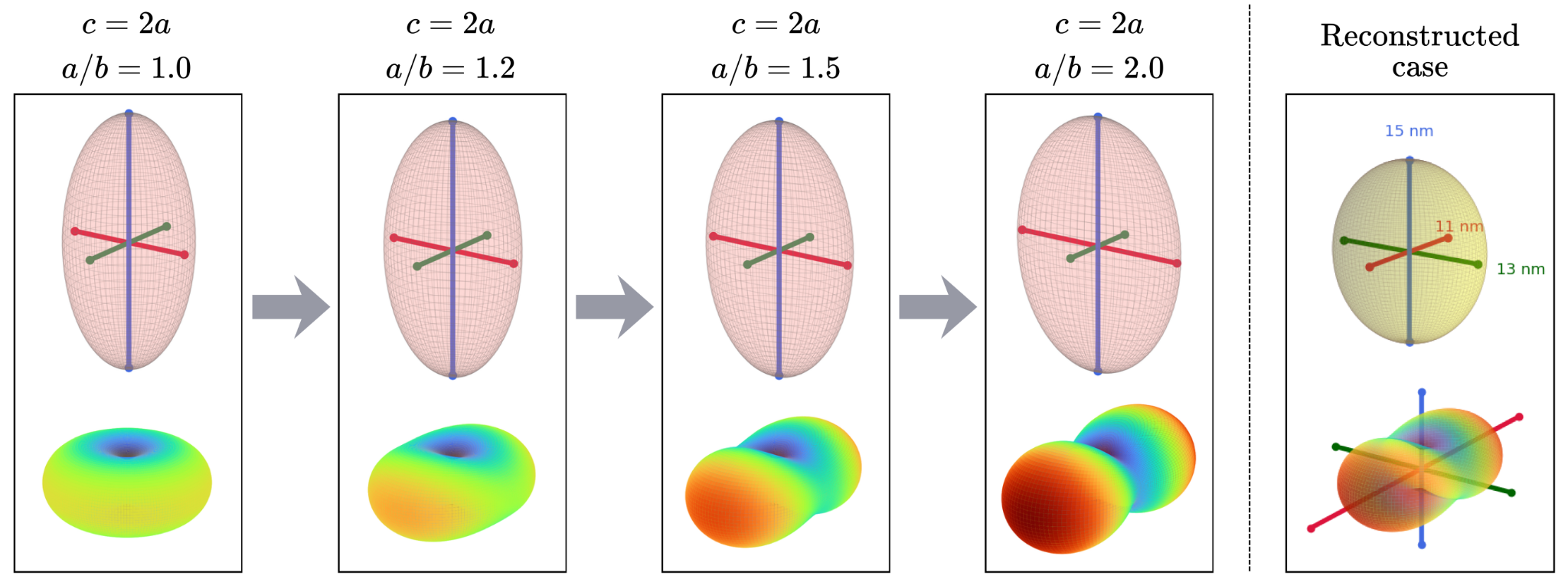}
    \caption{Evolution of the effective shape-anisotropy energy landscape with increasing particle triaxiality. The four left panels illustrate a systematic progression from the axially symmetric case to increasingly triaxial geometries, obtained by varying the ratio \(a/b\) from 1.0 to 2.0 while keeping \(c=2a\). For each geometry, the particle shape is shown above and the corresponding shape-anisotropy energy surface below. The rightmost panel, separated by the dashed line, shows the representative triaxial particle obtained from the inverse TEM reconstruction, with principal-axis dimensions of \(15\times13\times11~\mathrm{nm}^3\), together with its corresponding shape-anisotropy energy surface.}
    \label{fig:triaxial_energy_surface}
\end{figure*}

Since \(K_c>K_b\), rotations of the magnetization towards the short axis are energetically more costly than rotations towards the intermediate axis. Thus, the triaxial geometry gives rise to an effective biaxial shape anisotropy, rather than to a purely uniaxial contribution. The progressive emergence of this biaxial character is illustrated in Figure~\ref{fig:triaxial_energy_surface}. Starting from the axially symmetric case, increasing the difference between the two transverse particle dimensions progressively breaks the rotational symmetry of the shape anisotropy. As a consequence, the corresponding anisotropy-energy surface evolves from an uniaxial landscape towards a clearly biaxial, two-lobed structure, reflecting the increasing energetic distinction between rotations towards the two transverse particle axes. For implementation in OOMMF, the same shape-anisotropy energy density was expressed as the sum of two uniaxial terms,

\begin{equation}
\varepsilon_{\mathrm{shape}}
=
K_b\left(1-m_a^2\right)
+
\left(K_c-K_b\right)m_c^2.
\end{equation}
The first term defines an effective easy axis along the longest particle axis, while the second term introduces an additional hard-axis contribution along the shortest axis. For the \(15\times13\times11~\mathrm{nm}^3\) particle considered here, the demagnetizing factors are approximately \cite{osborn1945demagnetizing}

\begin{equation}
N_a\simeq0.2745,\qquad
N_b\simeq0.3274,\qquad
N_c\simeq0.3981.
\end{equation}

These values lead to

\begin{equation}
K_b\simeq7.7\times10^3~\mathrm{J\,m^{-3}},
\qquad
K_c\simeq1.8\times10^4~\mathrm{J\,m^{-3}}.
\end{equation}
The rightmost panel of Figure~\ref{fig:triaxial_energy_surface} illustrates the representative mean triaxial geometry obtained from the inverse TEM reconstruction, with principal-axis dimensions of \(15\times13\times11~\mathrm{nm}^3\), together with its corresponding shape-anisotropy energy surface. The two equivalent energy minima lie along the longest particle axis, while rotations towards the intermediate and short axes involve progressively larger energy penalties.

The second row of Figure~\ref{fig:micro_macro_combined} presents the energy density as a function of the applied-field polar angle \(\theta_H\) for the three azimuthal planes introduced above. The energy minimum occurs along the longest particle axis, whereas rotations towards the intermediate and short axes progressively increase the shape-anisotropy energy. The different energy barriers obtained for \(\phi_H=0^\circ\), \(45^\circ\), and \(90^\circ\) reflect the biaxial character of the triaxial particle geometry.

The total macrospin energy density was then written as

\begin{equation}
\varepsilon_{\mathrm{macro}}
=
\varepsilon_{\mathrm{Z}}
+
\varepsilon_{\mathrm{cub}}
+
\varepsilon_{\mathrm{shape}},
\end{equation}
where \(\varepsilon_{\mathrm{Z}}\) is the Zeeman contribution, \(\varepsilon_{\mathrm{cub}}\) is the cubic magnetocrystalline anisotropy energy density, and \(\varepsilon_{\mathrm{shape}}\) is the effective biaxial shape-anisotropy contribution derived from the triaxial geometry.

\subsubsection{Comparison between micromagnetic and macrospin simulations}

To compare the full micromagnetic simulations with the effective macrospin model, hysteresis loops were calculated for the same three azimuthal planes, \(\phi_H=0^\circ\), \(45^\circ\), and \(90^\circ\), while varying the polar angle \(\theta_H\) between the long particle axis and the corresponding transverse direction.

Representative individual hysteresis loops for \(\theta_H=0^\circ\), \(30^\circ\), \(60^\circ\), and \(90^\circ\) are shown in the two rows immediately below the energy profiles in Figure~\ref{fig:micro_macro_combined}.  Overall, the macrospin model reproduces the micromagnetic hysteresis loops closely over the investigated angular range. To obtain a response representative of an orientational distribution in the polar angle, the hysteresis loops were subsequently averaged using the \(\sin\theta_H\) solid-angle weighting. For each value of the azimuthal angle \(\phi_H\), the normalized averaged magnetization was calculated as

\begin{equation}
\overline{M}(H) =
\frac{
\int_0^{\pi/2}
M(H,\theta_H)\sin(\theta_H)\,d\theta_H
}{
\int_0^{\pi/2}
\sin(\theta_H)\,d\theta_H
}.
\label{eq:average_magnetization}
\end{equation}
The same averaging procedure was applied to both the micromagnetic and macrospin calculations. The bottom row of Figure~\ref{fig:micro_macro_combined} shows the averaged hysteresis loops obtained for the three azimuthal directions considered. The agreement between the micromagnetic and macrospin results is very good under the conditions considered, indicating that the magnetization reversal of the \(15\times13\times11~\mathrm{nm}^3\) triaxial NP remains essentially coherent. Therefore, the effective macrospin model, including cubic magnetocrystalline anisotropy and the biaxial shape-anisotropy contribution derived from the demagnetizing factors, provides an accurate reduced description of the micromagnetic response.

\section*{Acknowledgments}

This work was supported by the Spanish Ministry of Science, Innovation and Universities (MICIU) and the State Research Agency (AEI, 10.13039/501100011033) through Grants No. PID2024-157172NB-I00, No. PID2025-171111OB-I00, No. CNS2023-144321, through a Ramón y Cajal Fellowship (No. RYC2020-029822-I) to D.S and a Juan de la Cierva Fellowship (No. JDC2024-053418-I) to Y.F.A., co-funded by the European Social Fund ("ESF Investing in your future"). The Severo Ochoa Centres of Excellence program through Grants CEX2023-001286-S and CEX2024-001445-S funded by MICIU/AEI/10.13039/501100011033 is also acknowledged. Funding was also received by the Generalitat de Catalunya through Grant No. 2021SGR0032; by the Xunta de Galicia through Grants No. ED431F 2022/005 and No. ED431B 2023/055. T.S.v.Z also acknowledges the funding support from ARAID. We also acknowledge the Centro de Supercomputación de Galicia (CESGA) for providing computing resources and the use of the Advanced Microscopy Laboratory (Universidad de Zaragoza), for access to their instrumentation and expertise.

\section*{Conflicts of Interest}

The authors declare no conflicts of interest.

\section*{Data Availability Statement}

The data that support the findings of this study are available from the corresponding authors upon reasonable request.




\bibliography{wileyNJD-Chicago-abbrev_DOIs}

\clearpage
\onecolumngrid
\begin{center}
\textbf{ Supporting Information}
\end{center}
\vspace{1cm}


\section*{TEM analysis of the cellular internalization dataset}

\begin{figure*}[!h]
    \centering
    \includegraphics[width=1.0\textwidth]{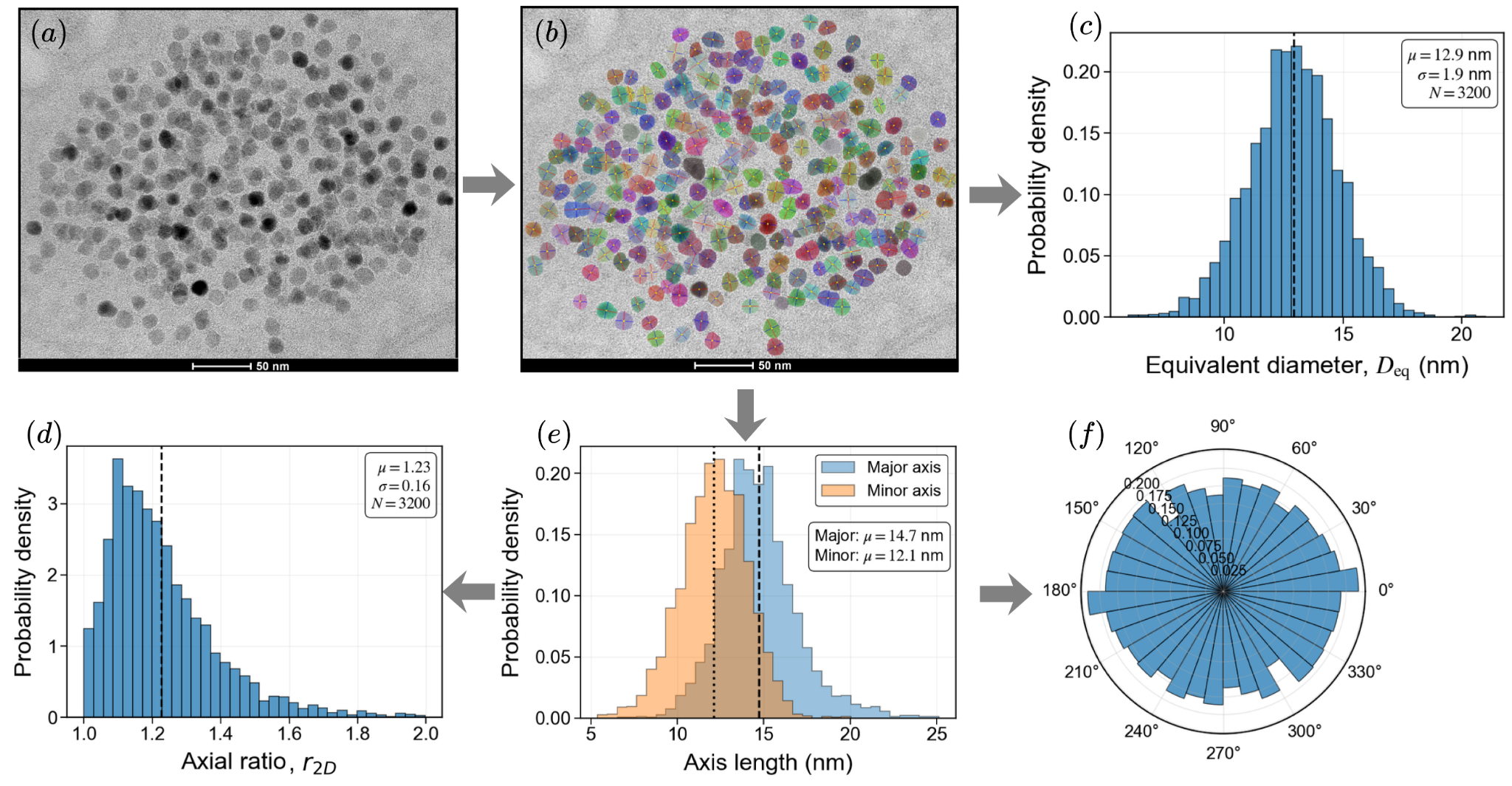}
    \caption{
TEM image analysis and projected morphological characterization of IONPs
observed after cellular internalization and embedding in resin.
(a) Representative TEM micrograph of the intracellular nanoparticle
population; scale bar: 50 nm.
(b) Same micrograph after CellposeSAM segmentation and PCA-based
morphological analysis, with the projected major and minor axes overlaid
on each detected particle; scale bar: 50 nm.
(c) Probability-density distribution of the projected equivalent diameter,
$D_{\mathrm{eq}}$, yielding a mean value of $\mu=12.9$ nm and a standard
deviation of $\sigma=1.9$ nm ($N=3200$).
(d) Probability-density distribution of the projected axial ratio,
$r_{2D}=a_{\mathrm{major}}/a_{\mathrm{minor}}$, yielding
$\mu=1.23$ and $\sigma=0.16$ ($N=3200$).
(e) Probability-density distributions of the projected major- and
minor-axis lengths, with mean values of $14.7$ and $12.1$ nm,
respectively. Vertical lines indicate the corresponding mean values.
(f) Polar distribution of the projected major-axis orientations in the
TEM image plane, showing no clear preferential in-plane orientation.
}
    \label{fig:TEM_cell_summary}
\end{figure*}

\clearpage

\section*{Particle-level comparison between experimental conditions}

\begin{figure*}[!h]
    \centering
    \includegraphics[width=1.0\textwidth]{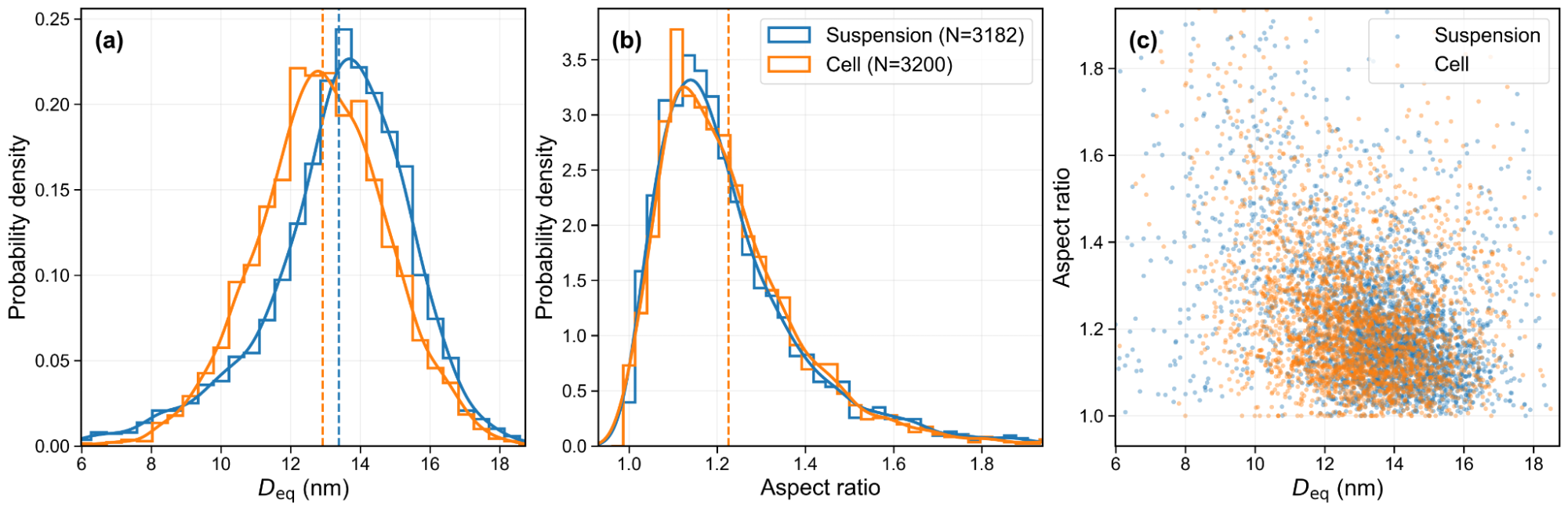}
    \caption{
    Particle-level comparison between particles deposited from suspension on TEM grids and particles measured inside cells. (a) probability-density distribution of the equivalent diameter \(d_{\mathrm{eq}}\). (b) probability-density distribution of the aspect ratio \(AR\). (c) relation between equivalent diameter and aspect ratio for individual particles.
    }
    \label{fig:particle_level_comparison}
\end{figure*}
\clearpage

\section*{Magnetic fluid hyperthermia simulations at 1 MHz}

\begin{figure}[!h]
    \centering
    \includegraphics[width=0.6\textwidth]{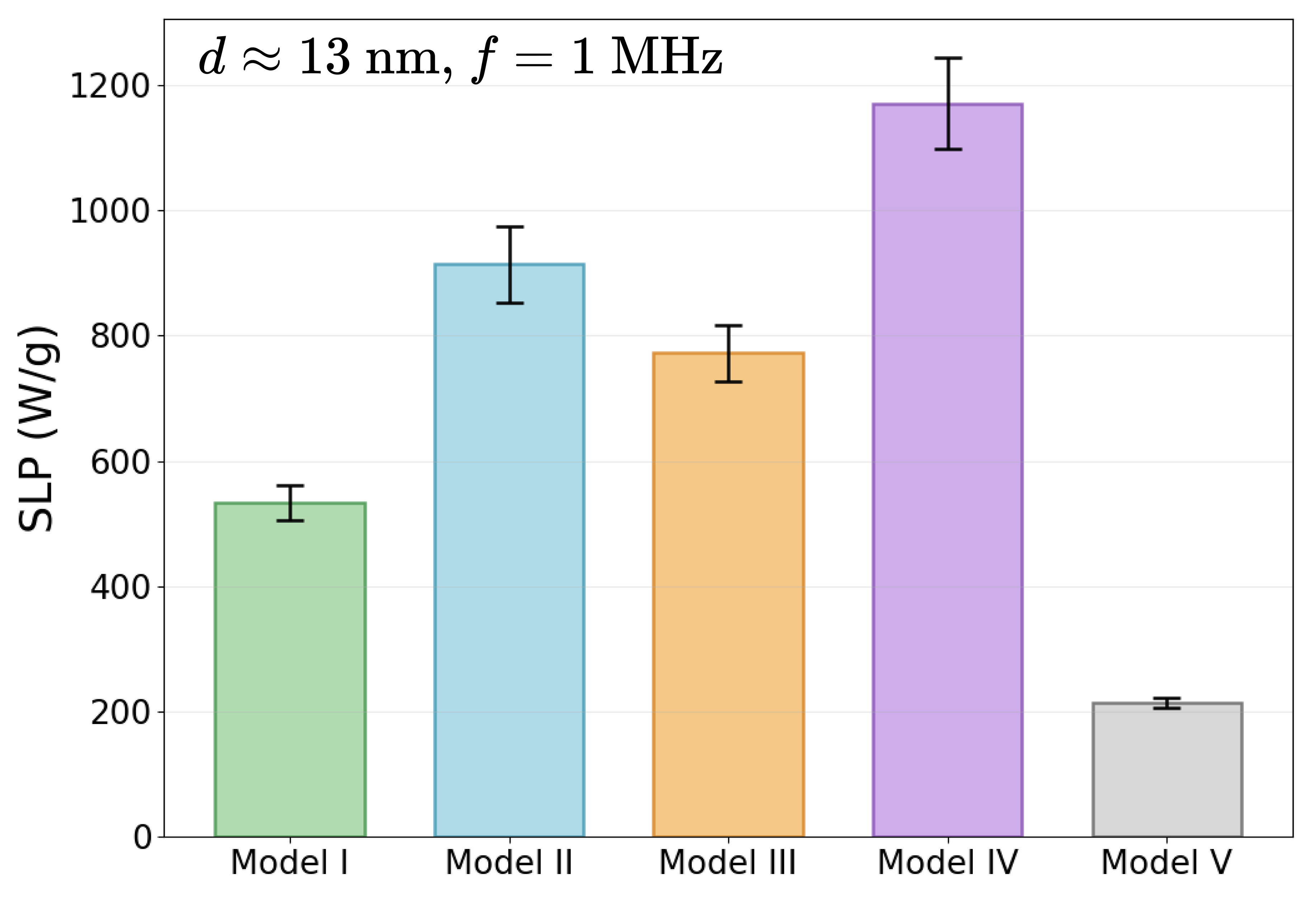}
    \caption{
    Specific loss power (SLP) predicted for nanoparticle ensembles generated according to the five 3D morphological models obtained from the inverse TEM reconstruction described in Sections~3.2--3.6. The particle-size distributions correspond to the reconstructed population centered around approximately \(d \approx 13\) nm. Results were obtained at \(f=1~\mathrm{MHz}\) and \(\mu_0H_{\max}=50~\mathrm{mT}\). Models I--IV correspond to prolate descriptions with different assumptions regarding size and shape polydispersity, whereas Model V represents a polydisperse population of triaxial ellipsoids. Error bars indicate the standard deviation obtained from the simulated ensemble.
    }
    \label{fig:SLP_1MHz}
\end{figure}

\end{document}